\documentclass[twocolumn,showpacs,superscriptaddress,preprintnumbers,amsmath,amssymb]{revtex4}

\usepackage{graphicx}
\usepackage{dcolumn}
\usepackage{bm}

\usepackage{color}
\usepackage{CJK}

\begin{document}



\title{Beam energy dependence of Hanbury-Brown-Twiss radii from a blast-wave model}

\author{S. Zhang}
\affiliation{Shanghai Institute of Applied Physics, Chinese Academy of Sciences, Shanghai 201800, China}
\author{Y. G. Ma\footnote{Author to whom all correspondence should be addressed: ygma@sinap.ac.cn}}
\affiliation{Shanghai Institute of Applied Physics, Chinese Academy of Sciences, Shanghai 201800, China}
\affiliation{University of Chinese Academy of Sciences, Beijing 100049, China}
\author{J. H. Chen}
\affiliation{Shanghai Institute of Applied Physics, Chinese Academy of Sciences, Shanghai 201800, China}
\author{C. Zhong}
\affiliation{Shanghai Institute of Applied Physics, Chinese Academy of Sciences, Shanghai 201800, China}

\date{\today}

\begin{abstract}
The beam energy dependence of correlation lengths (the Hanbury-Brown-Twiss radii) is calculated by using a blast-wave model and the results are comparable with those from RHIC-STAR beam energy scan data as well as the LHC-ALICE measurements. A set of parameter for the blast-wave model as a function of beam energy under study are obtained by fit to the HBT radii at each energy point. The transverse momentum dependence of HBT radii is presented with the extracted parameters for Au + Au collision at $\sqrt{s_{NN}} = $ 200 GeV and for Pb+Pb collisions at 2.76 TeV. From our study one can learn that particle emission duration can not be ignored while calculating the HBT radii with the same parameters. And tuning kinetic freeze-out temperature in a range will result in system lifetime changing in the reverse direction as it is found in RHIC-STAR experiment measurements.
\end{abstract}

\maketitle

\section{Introduction}

The Quark-Gluon-Plasma (QGP) predicted by quantum chromodynamics (QCD)~\cite{QCD-QGP} can be formed in relativistic heavy-ion collisions. It is believed that this kind of new state of matter is produced in the early stage of central Au + Au collisions at the top energy in the Relativistic Heavy-Ion Collider (RHIC) at Brookhaven National Laboratory~\cite{RHICWithePaper}. It was concluded that the hot-dense matter is a strongly interacting partonic matter named as sQGP under extreme temperature and energy density with sufficient experimental evidences~\cite{RHIC-SQGP,RHIC-SQGP2,RHIC-SQGP3,RHIC-SQGP4,RHIC-SQGP5}. Recently, many results in Pb + Pb and p+Pb collisions at $\sqrt{s_{NN}}$ = 2.76 TeV in the Large Hadron Collider (LHC) were also reported for exploring properties of the hot-dense quark-gluon matter ~\cite{ALICE-chDen,ALICE-chCDepen,ALICE-chRAA,ALICE-pionSpectra}.

Mapping the QCD phase diagram and locating the phase boundary and possible critical end point becomes hot topic in the field~\cite{Itoch-QaurkStars,QCD-QGP,Alford,Costa,Zong}. The properties inherited from QGP will imprint signal on observables which can reflect phase transition information. The geometry of the system shall undergo phase space evolution from QGP stage to hadron kinetic freeze-out stage, which can be considered as an observable that is sensitive to the equation of state~\cite{HBT-BES-Th, HBT-BES-EXP-STAR}. Hanbury-Brown-Twiss (HBT) technique invented for measuring sizes of nearby stars~\cite{HBT-STARSize} was extended to particle physics~\cite{HBT-Particle} and heavy-ion collisions~\cite{Koonin,Pratt,HBT-HeavyIon,Boal,HBT-UHeinz,Ma,Hu,ZhangZQ}. The HBT technique can also be applied to extract the precise space-time properties from particle emission region at kinetic freeze-out stage in heavy-ion collisions. Furthermore this technique has been evolved to  search for new particles and to measure particle interactions~\cite{STAR-Lambda-Lambda, STAR-pLambda, STAR-antiProtonInter}.

Experimental results on HBT study in high energy nuclear reaction were reported by STAR~\cite{HBT-STAR-200GeV} and PHENIX~\cite{HBT-PHENIX-200GeV} at RHIC top energy in Au+Au collisions, as well as by ALICE~\cite{HBT-ALICE} at $\sqrt{s_{NN}} = $ 2.76 TeV in Pb+Pb collisions. Recently STAR and PHENIX Collaborations have also presented beam energy dependence of HBT radii~\cite{HBT-BES-EXP-STAR, HBT-BES-EXP-PHENIX} and a non-monotonic changing behaviour for the square difference between outward radius and sideward radius ($R_{out}^2-R_{side}^2$) with increase of beam energy was found. This behaviour could be sensitive to equation of state and was considered as a probe related to the critical end point of QGP phase transition~\cite{HBT-BES-Th}. A finite-size scaling (FSS) analysis of experimental data was performed in Ref.~\cite{HBT-BES-Th} and the analysis suggested that a second order phase transition was taken place with a critical end point located at a chemical freeze-out temperature of $\sim$ 165 MeV and a baryon chemical potential of $\sim$ 95 MeV.

In this paper we present beam energy dependence of HBT radii calculated from a blast-wave model. Firstly, experimental data of HBT radii from RHIC-STAR and LHC-ALICE are fitted and parameters for the blast-wave model are configured as a function of beam energy. The transverse momentum dependence of HBT radii are calculated at RHIC top energy and LHC energy with these parameters. From the results, it was found that particle emission duration is important for calculating transverse momentum dependence of HBT radii and changing of kinetic freeze-out temperature will result in system lifetime changing in reverse direction as that in the RHIC-STAR experimental analysis~\cite{HBT-BES-EXP-STAR}.

The paper  is organised as following. In Sec. II,  blast-wave model and HBT correlation function are briefly introduced. Some kinetic parameters are presented as a function of beam energy.  Section III presents energy dependence of extracted HBT radii with various kinetic temperature, system lifetime and particle emission duration etc. Transverse momentum dependence of HBT radii is discussed in Section IV. Finally Section V gives the summary.

\section{Blast-wave model and HBT correlation function}

The particle emission function $S(x,p)$ in heavy-ion collisions used in this study is similar as in reference~\cite{BLWave-Fabrice},
\begin{widetext}
\begin{eqnarray}
S(x,p) & = & m_T\cosh(\eta-Y)\Omega(r,\phi_s)e^{-(\tau-\tau_0)^2/2\Delta\tau^{2}}\frac{1}{e^{K\cdot u/T_{kin}}\pm1}\nonumber\\
 & = & m_T\cosh(\eta-Y)\Omega(r,\phi_s)e^{-(\tau-\tau_0)^2/2\Delta\tau^{2}}\sum^{\infty}_{n=1}(\mp)^{n+1}e^{-K\cdot u/T_{kin}}\nonumber\\
 & \simeq & m_T\cosh(\eta-Y)\Omega(r,\phi_s)e^{-(\tau-\tau_0)^2/2\Delta\tau^{2}}e^{-K\cdot u/T_{kin}}.
\label{eq:EmissionFun}
\end{eqnarray}
\end{widetext}
In cylindrical coordinates, source moving four-velocity and momentum can be written respectively as,
\begin{eqnarray}
u_{\mu}(x) = & (\cosh\eta\cosh\rho(r,\phi_s),\sinh\rho(r,\phi_s)\cos\phi_b,\nonumber\\
                     & \sinh\rho(r,\phi_s)\sin\phi_b,\sinh\eta\cosh\phi_b),
\label{eq:Velocity}
\end{eqnarray}
and
\begin{eqnarray}
K_{\mu} = & (m_T\cosh Y, p_T\cos\phi_p, p_T\sin\phi_p, m_T\sinh Y).
\label{eq:Momentum}
\end{eqnarray}
And the flow rapidity is given by,
\begin{equation}
\rho(r,\phi_s) = \tilde{r}[\rho_0+\rho_2\cos(2\phi_b)],
\label{eq:FlowRap}
\end{equation}
here the normalized elliptical radius,
\begin{equation}
\tilde{r} \equiv \sqrt{\frac{[r\cos\phi_s]^2}{R_x^2}+\frac{[r\sin\phi_s]^2}{R_y^2}},
\label{eq:Tilde_r}
\end{equation}
with
\begin{equation}
\tan\phi_s = \left(\frac{R_y}{R_x}\right)^2\tan\phi_b.
\label{eq:phis_phib}
\end{equation}

In equation Eq.~(\ref{eq:EmissionFun}), spatial weighting of source elements is selected as a simple pattern~\cite{BLWave-Fabrice},
\begin{eqnarray}
\Omega(r,\phi_s) = \left\{ \begin{array}{ll}
1 &, \tilde{r}<1\\
0 &, \tilde{r}>1
\end{array}
\right    ..
\label{eq:Omega}
\end{eqnarray}

\begin{figure}
\includegraphics[width=8.2cm]{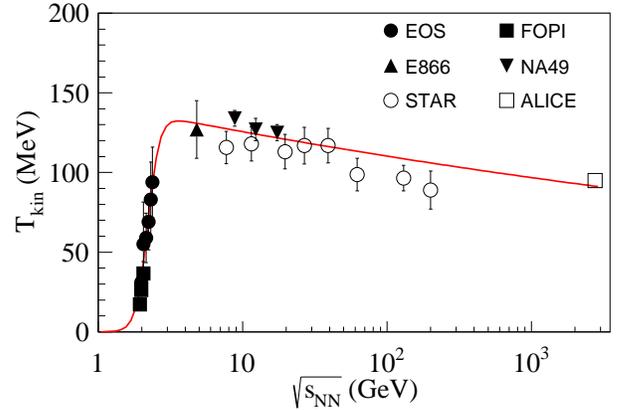}
\caption{\label{fig:Tkin} (Color online)
Kinetic freeze-out temperature as a function of  centre-of-mass energy $\sqrt{s_{NN}}$. Data are from~\cite{STARSYS-Spectra,STAR-Kumar-BES}.
}
\end{figure}

\begin{figure}
\includegraphics[width=8.2cm]{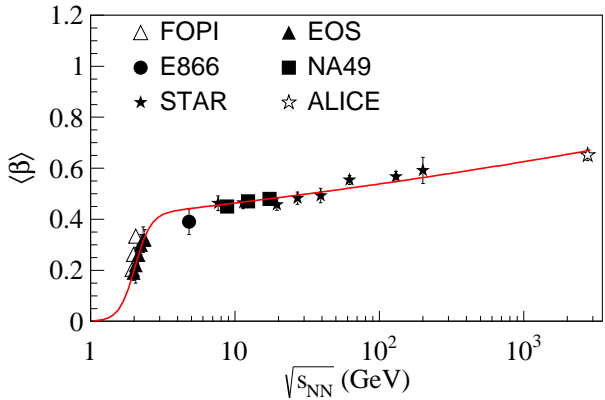}
\caption{\label{fig:beta} (Color online)
The averaged radial flow $\langle\beta\rangle$ as a function of  center-of-mass energy $\sqrt{s_{NN}}$. Data are from~\cite{STARSYS-Spectra,STAR-Kumar-BES}.
}
\end{figure}

Here are the main parameters in this model, the kinetic freeze-out temperature $T_{kin}$, the radial flow parameter $\rho_0$, the "elliptic flow parameter" $\rho_2$ which controls second-order oscillation of transverse rapidity by the relation as in Eq.~(\ref{eq:phis_phib}), the system lifetime $\tau_0$ and the particle emission duration $\Delta\tau$, $R_x$ and $R_y$ related to system size and space asymmetry. In this calculation we assume that the system is in most central heavy-ion collisions and thus set the $R_x = R_y = R_0$, $\rho_2 = 0$. In experimental measurement, hadron spectra can be fitted by the blast-wave model with integrating the emission function except the $p_T$ and $Y$. The kinetic freeze-out temperature $T_{kin}$ and the averaged radial flow $\langle\beta\rangle$ was extracted from the fit. For detail technique information, one may refer to~\cite{STARSYS-Spectra}. The averaged radial flow is related to the flow rapidity $\rho = \tanh^{-1}\beta$, from which the radial flow parameter $\rho_0$ is calculated. Figure~\ref{fig:Tkin} and~\ref{fig:beta}  present the measured $T_{kin}$ and $\langle\beta\rangle$ at a wide beam energy range respectively. The data come from~\cite{STARSYS-Spectra,STAR-Kumar-BES}. The kinetic freeze-out temperature $T_{kin}$ and the averaged radial flow $\langle\beta\rangle$ can be parametrised as a function of $\sqrt{s_{NN}}$ by empirical formula,
\begin{widetext}
\begin{eqnarray}
      T_{kin} & = & T_{lim}\frac{1}{(1+\exp(8.559-\ln(\sqrt{s_{NN}})/0.093))/(\sqrt{s_{NN}}^{0.057}/0.846)}\nonumber\\
\langle\beta\rangle & = & \beta_{lim}\frac{1}{(1+\exp(5.666-\ln(\sqrt{s_{NN}})/0.124))\sqrt{s_{NN}}^{0.065})},
\label{eq:TBeta_sNN}
\end{eqnarray}
\end{widetext}
where $T_{lim}$ = 169.171 MeV and $\beta_{lim}$ = 0.399. And then free parameters in the blast-wave model will be the $R_0$, the $\tau_0$ and the $\Delta\tau$, which are all related to expanding characters of the collision system. And it will be determined by the HBT correlation calculation which will be discussed below in detail.

In our previous works, the blast-wave model was coupled with thermal equilibrium model to describe the hadron production and its spectra with a range of thermal parameters~\cite{SZhang-LHC}, and with coalescence mechanism to calculate the light nuclei production and to predict the di-baryons  production rate~\cite{BW-XueLiang,BW-Neha}.  In addition, the DRAGON model~\cite{DRAGON} and the THERMINATOR2~\cite{THERMINATOR}
model have also been developed as event generator to study the phase-space distribution of  hadrons at freeze-out stage. It is also successfully applied in experimental data analysis~\cite{STARSYS-Spectra,STAR-Kumar-BES} to extract the kinetic freeze-out properties and to provide the phase-space distribution to calculate the HBT correlation in theory~\cite{BLWave-Fabrice,HBT-UAW}. 

The identical two particle HBT correlation function can be written as~\cite{HBT-UAW,HBT-UHeinz},
\begin{eqnarray}
C(\vec{K},\vec{q}) = 1+
\left|\frac{\int d^4xe^{i\vec{q}\cdot(\vec{x}-\vec{\beta}t)}S(x,K)}{\int d^4xS(x,K)}\right|,
\label{eq:HBTCorrFun}
\end{eqnarray}
here $K$ is average momentum for the two particles, $K=\frac{1}{2}(p_1+p_2)$, $q$ denotes relative momentum between two particles, $q=p_1-p_2$, and $\vec{\beta}=\vec{K}/\vec{K_0}$. From Refs.~\cite{HBT-UAW,BLWave-Fabrice, HBT-oslSys}, the ``out-side-long" coordinates system is used in this calculation, in which the long direction $R_{long}$  is parallel to the beam, the sideward direction $R_{side}$  is perpendicular to the beam and total pair momentum, and the outward direction  $R_{out}$  is perpendicular to the long and sideward directions. After expanding angular dependence of $C(K,q)$ in a harmonic series with the ``out-side-long" coordinates system, the HBT radii can be written as~\cite{HBT-UAW,BLWave-Fabrice},
\begin{widetext}
\begin{eqnarray}
R_{side}^2&=&\frac{1}{2}(\langle\tilde{x}^2\rangle+\langle\tilde{y}^2\rangle)-\frac{1}{2}(\langle\tilde{x}^2\rangle-\langle\tilde{y}^2\rangle)\cos(2\phi_p)-\langle\tilde{x}\tilde{y}\rangle\sin(2\phi_p),\nonumber\\
R_{out}^2&=&\frac{1}{2}(\langle\tilde{x}^2\rangle+\langle\tilde{y}^2\rangle)+\frac{1}{2}(\langle\tilde{x}^2\rangle-\langle\tilde{y}^2\rangle)\cos(2\phi_p)+\langle\tilde{x}\tilde{y}\rangle\sin(2\phi_p)-2\beta_T(\langle\tilde{t}\tilde{x}\rangle\cos\phi_p+\langle\tilde{t}\tilde{y}\rangle\sin\phi_p)+\beta_T^2\langle\tilde{t}^2\rangle,\nonumber\\
R_{long}^2&=&\langle\tilde{z}^2\rangle-2\beta_l\langle\tilde{t}\tilde{z}\rangle+\beta_l^2\langle{\tilde{t}}^2\rangle,
\label{eq:HBT-Radii}
\end{eqnarray}
\end{widetext}
where
\begin{eqnarray}
\langle f(x)\rangle(K)&\equiv&\frac{\int d^4xf(x)S(x,K)}{\int d^4xS(x,K)},\nonumber\\
           \tilde{x}^{\mu}&\equiv&x^{\mu}-\langle\tilde{x}^{\mu}\rangle(K).
\end{eqnarray}

In the calculation, observables are related to integrals of emission function~(\ref{eq:EmissionFun}) over phase space $d^4x=dxdydzdt=\tau d\tau d\eta rdrd\phi_s$, weighted with some quantities $B(x,K)$. If $B(x,K)=B'(r,\phi_s,K)\tau^i\sinh^j\eta\cosh^k\eta$, then the integrals can be written as in~\cite{BLWave-Fabrice},
\begin{eqnarray}
\int^{2\pi}_0d\phi_s\int^{\infty}_0rdr\int^{\infty}_{-\infty}d\eta\int^{\infty}_{-\infty}\tau d\tau S(x,K)B(xK)\nonumber\\
=m_TH_i\{B'\}_{j,k}(K),
\end{eqnarray}
and some useful integrals,
\begin{eqnarray}
H_i&\equiv&\int^{\infty}_{-\infty}d\tau\tau^{i+1}e^{-(\tau-\tau_0)^2/2\Delta\tau^2},\nonumber\\
G_{j,k}(x,K)&\equiv&\int^{\infty}_{-\infty}d\eta e{-\beta\cosh\eta}\sinh^j\eta\cosh^{k+1}\eta,\nonumber\\
\{B'\}_{j,k}(K)&\equiv&\int^{2\pi}_0d\phi_s\int^{\infty}_0rdrG_{j,K}(x,K)B'(x,K)\nonumber\\
&\times&e^{\alpha\cos(\phi_b-\phi_p)}\Omega(r,\phi_s),
\end{eqnarray}
where we define,
\begin{eqnarray}
\alpha&\equiv&\frac{p_T}{T}\sinh\rho(r,\phi_s),\nonumber\\
\beta&\equiv&\frac{m_T}{T}\cosh\rho(r,\phi_s).
\end{eqnarray}

Retie\`ere and Lisa~\cite{BLWave-Fabrice} have provided a systematic analysis of parameter range for the blast-wave model and investigated the $p_T$ spectra, the collective flow, and the HBT correlation of hadrons produced in heavy-ion collisions. In this calculation we will use the algorithm developed in Ref.~\cite{BLWave-Fabrice,HBT-UAW} to study the energy and transverse momentum dependence of pion HBT correlation radii. Based on the discussion above, the free parameters will be $R_0$, $\tau_0$ and $\Delta\tau$ which can be determined by fitting experimental data by Eq.~(\ref{eq:HBT-Radii}). Before the study of energy dependence on HBT radii, we calculated pion's spectra by using this algorithm in the blast-wave model,
\begin{eqnarray}
\frac{dN}{p_Tdp_T} = \int d\phi_p\int d^4xS(x,K).
\end{eqnarray}

\begin{figure}
\includegraphics[width=8.2cm]{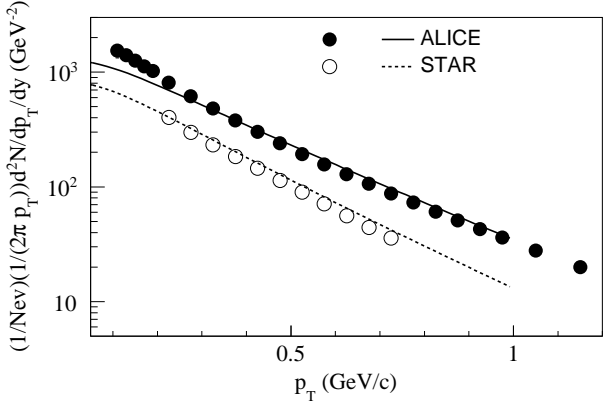}
\caption{\label{fig:pionSpectra} 
Comparison of pion's spectra from blast-wave model (lines) and the data in central Au+Au collisions at $\sqrt{s_{NN}}$ = 200 GeV~\cite{STAR-pionSpectra} and the data in central Pb+Pb collisions at $\sqrt{s_{NN}}$ = 2.76 TeV ~\cite{ALICE-pionSpectra}.
}
\end{figure}

Figure~\ref{fig:pionSpectra} presents pion's spectra which are comparable with experimental data from STAR at $\sqrt{s_{NN}}$ = 200 GeV in central Au+Au collisions~\cite{STAR-pionSpectra} and ALICE at $\sqrt{s_{NN}}$ = 2.76 TeV in central Pb+Pb collisions~\cite{ALICE-pionSpectra}, respectively. 

\section{energy dependence of HBT radii}
The parameters are configured as following. The kinetic freeze-out temperature $T_{kin}$ and the averaged radial flow $\langle\beta\rangle$ are from Eq.~(\ref{eq:TBeta_sNN}) as a function of $\sqrt{s_{NN}}$, but in some case  $T_{kin}$ are fixed to 90, 100 and 120 MeV for comparison. In numerical calculation, the particle emission duration $\Delta\tau$ is set to zero and in another case the energy dependence of $\Delta\tau$  will be extracted by fit on the data. The $R_0$ will also be given by fit the data at each energy point. The experimental results of HBT radii are taken from the STAR and the ALICE collaborations~\cite{HBT-BES-EXP-STAR,HBT-ALICE} at centre-of-mass energy $\sqrt{s_{NN}}$ points, 7.7, 11.5, 19.6, 27, 39, 62.4, 200 and 2760 GeV. The difference between calculated radii results and the experimental data should reach a minimum value ($\delta_s, \delta_o, \delta_l$) for each energy point,
\begin{eqnarray}
R_{side}(\text{th})-R_{side}(\text{exp})&=&\delta_s,\nonumber\\
R_{out}(\text{th})-R_{out}(\text{exp})&=&\delta_o,\nonumber\\
R_{long}(\text{th})-R_{long}(\text{exp})&=&\delta_l.\nonumber
\end{eqnarray}
Actually from Eq.(~\ref{eq:HBT-Radii}) and the algorithm in~\cite{BLWave-Fabrice,HBT-UAW}, one can find the HBT radii parameter dependence as following:
\begin{eqnarray}
R_{side}^2&=&R_{side}^2(T_{kin},\rho_0,R_0),\nonumber\\
R_{out}^2&=&R_{out}^2(T_{kin},\rho_0,R_0,\tau_0,\Delta\tau),\nonumber\\
R_{long}^2&=&R_{long}^2(T_{kin},\rho_0,R_0,\tau_0,\Delta\tau).
\end{eqnarray}
So $R_0$ can be determined directly by fit on $R_{side}^2$. And the $\tau_0$, $\Delta\tau$ can be extract by fit on $R_{out}^2$ and $R_{long}^2$ simultaneously. We then learnt that the difference of $R_{out}^2-R_{side}^2$ not only depends on the system lifetime $\tau_0$ but also on the particle emission duration $\Delta\tau$.

\begin{figure*}
\includegraphics[width=15cm]{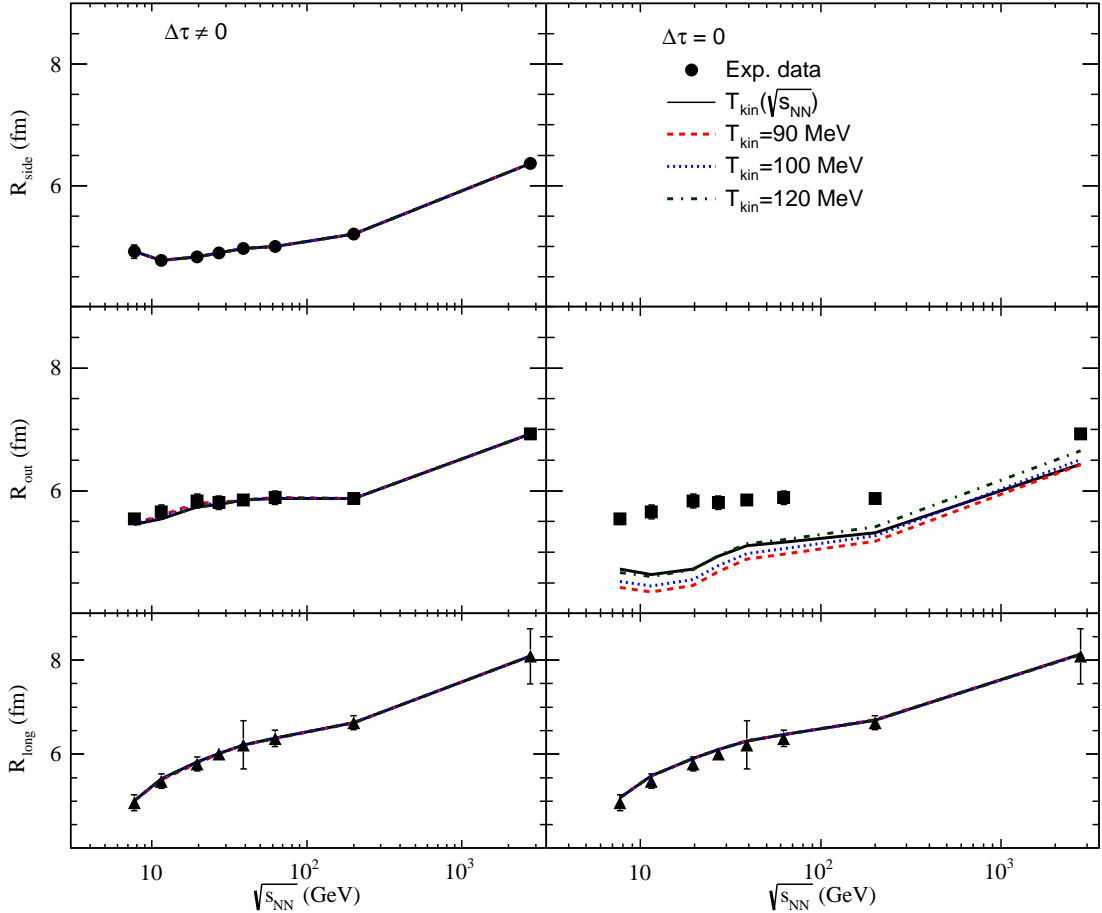}
\caption{\label{fig:Rosl} (Color online)The extracted values of 
$R_{side}$, $R_{out}$ and $R_{long}$ as a function of center-of-mass energy $\sqrt{s_{NN}}$. Different type of lines represent different kinetic temperature parametrisation. The solid points are experimental data  from~\cite{HBT-BES-EXP-STAR,HBT-ALICE}. Left panels are results with finite particle emission duration ($\Delta \tau$) and right panels for the cases of $\Delta \tau$=0.
}
\end{figure*}

 \begin{figure}
\includegraphics[width=8.2cm]{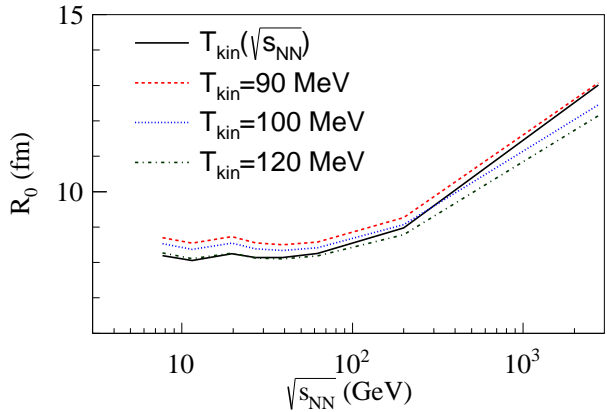}
\caption{\label{fig:R0} (Color online)
$R_0$ as a function of  center-of-mass energy $\sqrt{s_{NN}}$ with different kinetic temperature parametrisation.
}
\end{figure}

Figure~\ref{fig:Rosl} presents our calculation on HBT radii for identical charged pion-pion correlation with the configured parameters. The HBT radii show an increasing trend with the increasing of centre-of-mass energy $\sqrt{s_{NN}}$. In the case of $\Delta\tau\neq 0$, the results can describe experimental data successfully. However, for $\Delta\tau$ = 0.0,  the $R_{out}$ cannot be fitted despite $R_{long}$ can be well matched by the calculation. Since $T_{kin}$ and $\rho_0$ are taken from experimental results, $R_{side}$ will only depend on parameter $R_0$, which reflects the system size where particles are emitted. Figure~\ref{fig:R0} displays the extracted $R_0$ as a function of $\sqrt{s_{NN}}$. It demonstrates a similar trend of energy dependence as $R_{side}$. With fixed temperature of $T_{kin}$ (90, 100, 120 MeV), it is found that a large $R_0$ is needed to fit the data while $T_{kin}$ sets to small value. This is consistent with evolution of the fireball created in heavy-ion collisions, where temperature becomes lower while system size increases.

\begin{figure*}
\includegraphics[width=15cm]{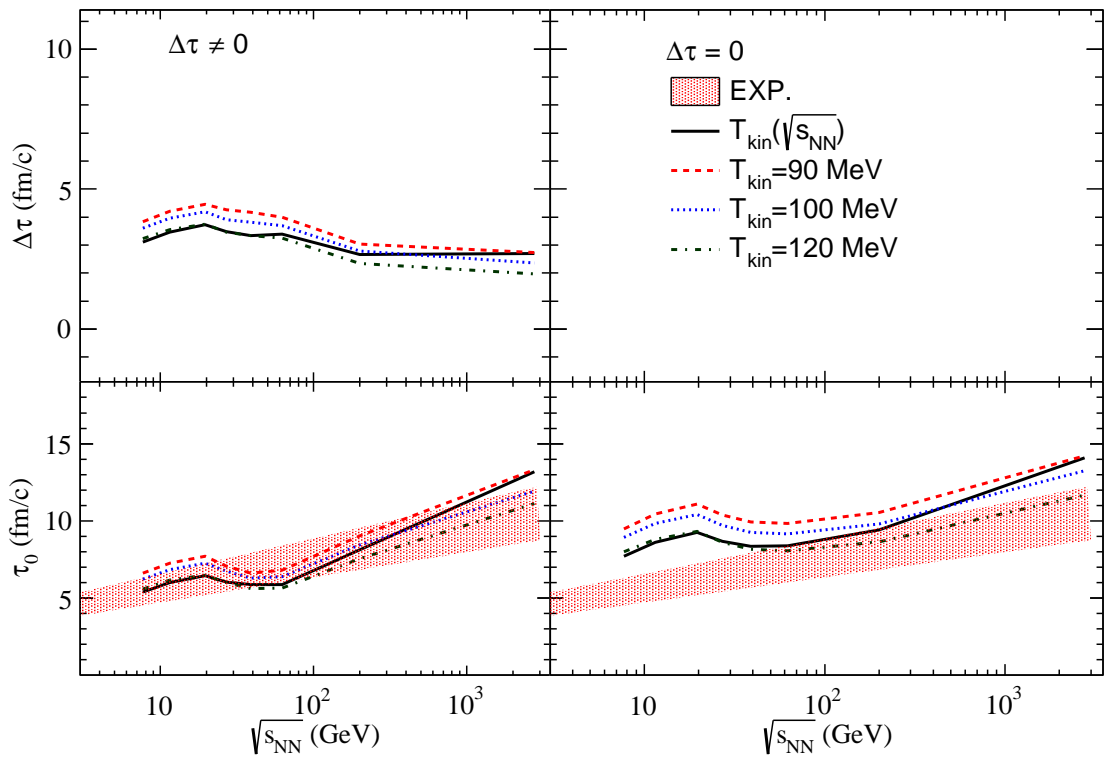}
\caption{\label{fig:tau0Deltatau} (Color online)
Same as Fig.\ref{fig:Rosl} but for $\tau_0$ and $\Delta\tau$. Experimental data (shadowed area) is taken  from~\cite{HBT-BES-EXP-STAR}.
}
\end{figure*}

$R_{out}$ and $R_{long}$ not only depend on $\tau_0$ but also on $\Delta\tau$. Figure~\ref{fig:tau0Deltatau} shows $\tau_0$ and $\Delta\tau$ as a function of  centre-of-mass energy $\sqrt{s_{NN}}$ from fit to the data. The $\Delta\tau$ slightly depends on the $\sqrt{s_{NN}}$.  From figure~\ref{fig:tau0Deltatau} one can see that $\tau_0$ generally increases with the increasing of $\sqrt{s_{NN}}$ in trends but there exists a minimum value at $\sqrt{s_{NN}}\sim$ 39 GeV. It may imply that the system in higher energy (such as at LHC) will undergo a longer time evolution than in lower energy before hadron rescattering ceases (the kinetic freeze-out status). With fixed temperature of $T_{kin}$ (90, 100, 120 MeV), the system lifetime $\tau_0$ and the particle emission duration $\Delta\tau$ are all in reverse order to the temperature $T_{kin}$. This suggests that a system expanding with a long lifetime and a broad duration will result in a lower temperature, which is consistent with the behaviour of $R_0$ as discussed above. We learnt that our results are comparable with the experimental results with $\Delta\tau\neq 0$. With the system lifetime and HBT radii calculation all taken into account, it can be concluded that the particle emission duration can not be ignored while fitting the HBT radii ($R_{side}$, $R_{out}$ and $R_{long}$) at the same time.

\begin{figure*}
\includegraphics[width=15cm]{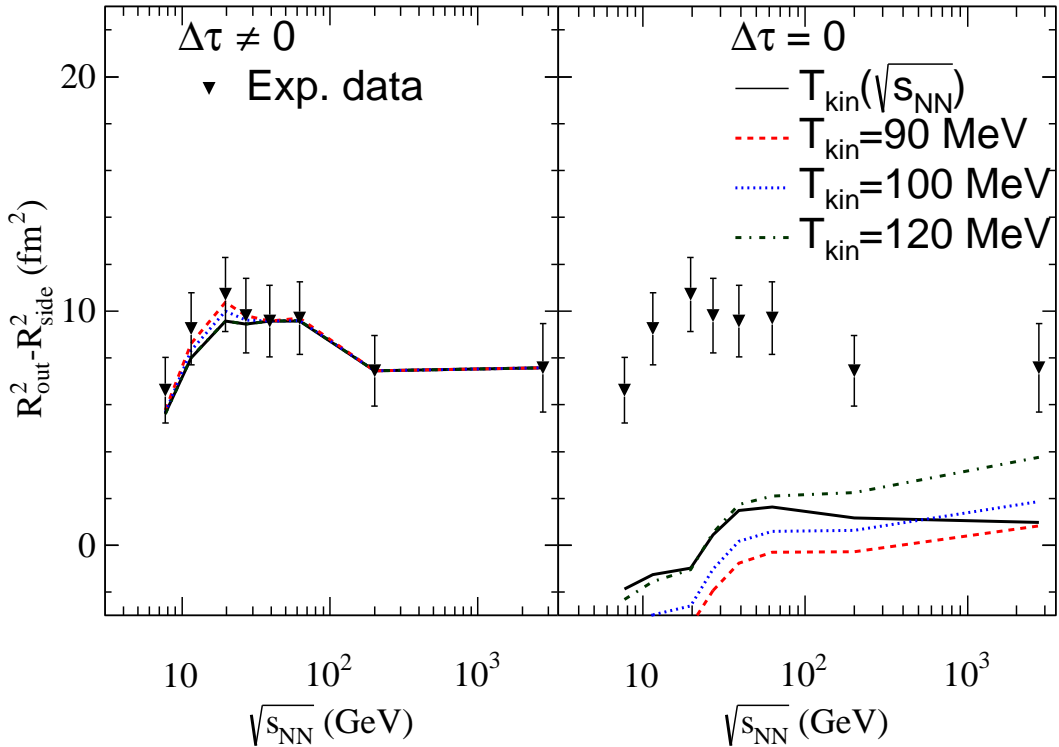}
\caption{\label{fig:RoRsDiff} (Color online)
Same as Fig.\ref{fig:Rosl} but for the $R_{out}^2-R_{side}^2$. Experiment data is taken from Ref.~\cite{HBT-BES-EXP-STAR}.
}
\end{figure*}

After $R_{out}$ and $R_{side}$ are all calculated, difference of $R_{out}^2-R_{side}^2$ as a function of  centre-of-mass energy $\sqrt{s_{NN}}$ can be obtained as shown in Figure~\ref{fig:RoRsDiff}. In the case of $\Delta\tau\neq 0$, the calculated results can describe the data very well. However, it is unsuccessful to fit the data with $\Delta\tau$=0 for the current parameter configuration. Energy dependence of the difference of $R_{out}^2-R_{side}^2$ demonstrates a non-monotonic increasing trend with the increasing of $\sqrt{s_{NN}}$. The peak of experimental results locates at $\sqrt{s_{NN}}\sim$17.3 GeV~\cite{HBT-BES-EXP-STAR} and the calculated results give a very similar behaviour for the peak emerging. And in reference~\cite{HBT-BES-Th}, the theoretical work proposes the critical end point (CEP) for deconfinement phase transition at $\sqrt{s_{NN}}$ = 47.5 GeV by applying FSS. Anyway other observables, such as elliptic flow and fluctuations, should be considered together and other basic theoretical calculations are awaiting for comparison, which contribute to locate the CEP and understand underlying physics around this energy region.

\section{transverse momentum dependence of HBT radii}

\begin{figure*}
\includegraphics[width=15cm]{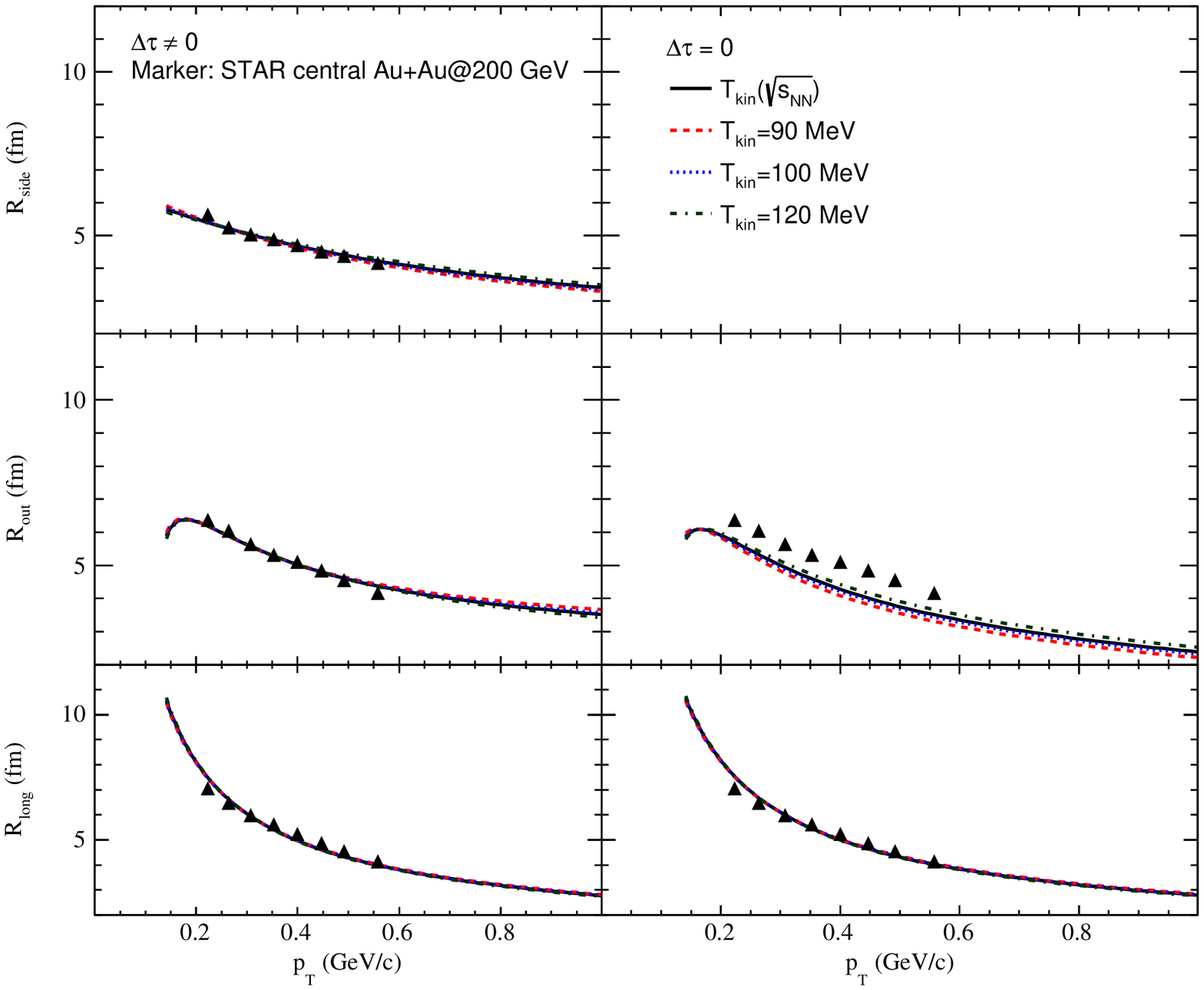}
\caption{\label{fig:RsolpTSTAR} (Color online)
The transverse momentum dependence of HBT radii in central Au+Au collisions at $\sqrt{s_{NN}}$=200 GeV. Experimental data is taken from Ref.~\cite{HBT-BES-EXP-STAR}.
}
\end{figure*}

\begin{figure*}
\includegraphics[width=15cm]{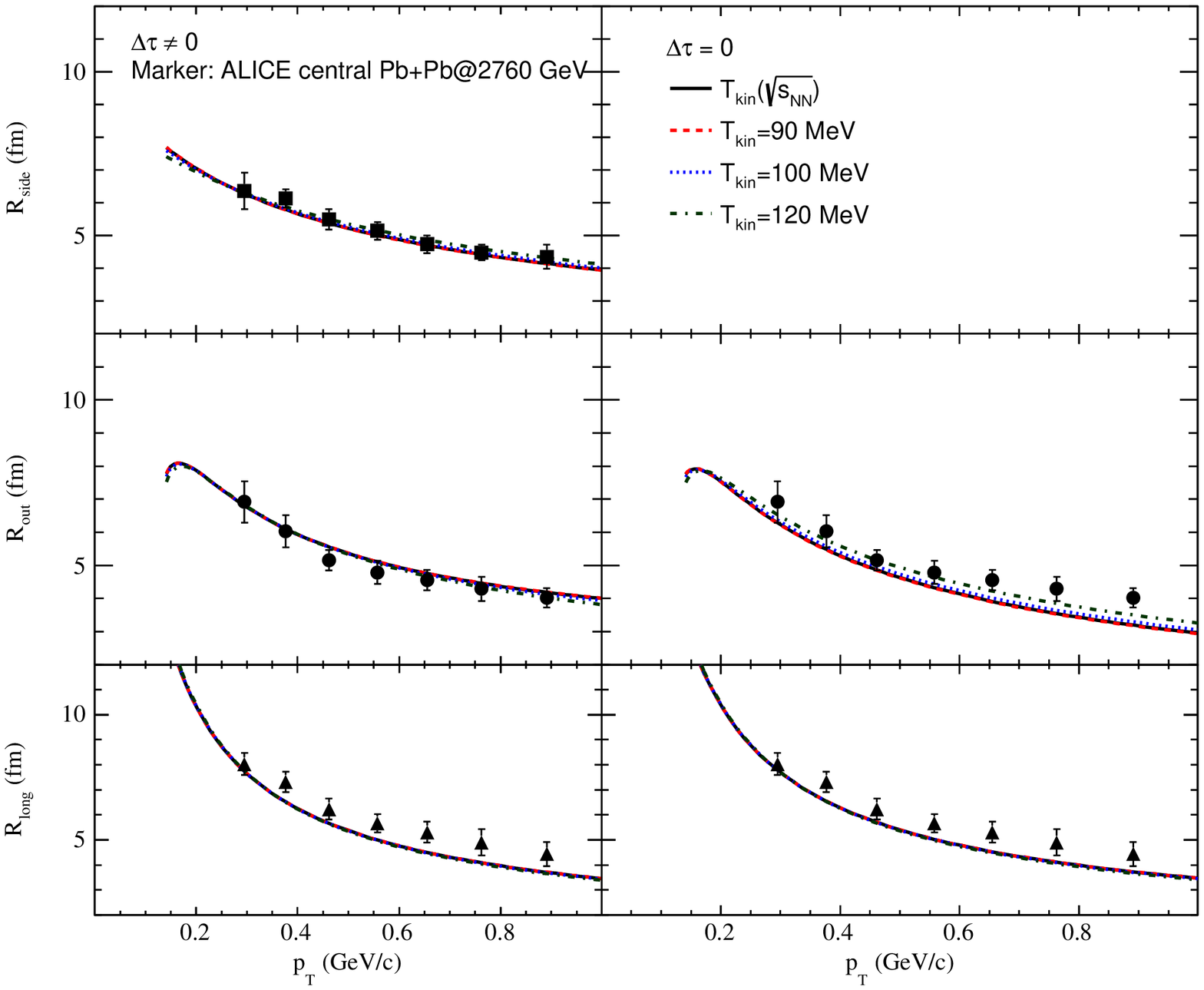}
\caption{\label{fig:RsolpTALICE} (Color online)
Same as Fig.~\ref{fig:RsolpTSTAR} but for the central Pb+Pb collisions at $\sqrt{s_{NN}}$=2760 GeV. Experimental data is taken from Ref.~\cite{HBT-ALICE}.
}
\end{figure*}

With the above parameter configuration, we also calculated the transverse momentum dependence of HBT radii at $\sqrt{s_{NN}}$=200 GeV and 2760 GeV in central heavy-ion collisions. Figure~\ref{fig:RsolpTSTAR} and~\ref{fig:RsolpTALICE} show the HBT radii as a function of transverse momentum in central Au+Au collisions at $\sqrt{s_{NN}}$=200 GeV and in central Pb+Pb collisions at $\sqrt{s_{NN}}$=2760 GeV, respectively. The experimental data is from Ref.~\cite{HBT-BES-EXP-STAR,HBT-ALICE}. $R_{side}$, $R_{out}$ and $R_{long}$ decrease with the increasing of transverse momentum $p_T$ as shown in figure~\ref{fig:RsolpTSTAR}, which indicates that high $p_T$ particles are emitted from near the centre of the fireball. It is found that the calculated results fit the STAR data in the case of $\Delta\tau\neq 0$ but fails to describe the $R_{out}$ with  $\Delta\tau$ = 0. The similar $p_T$ dependence trend is found in central Pb+Pb collisions at $\sqrt{s_{NN}}$ = 2760 GeV as shown in figure~\ref{fig:RsolpTALICE}. In the $\Delta\tau\neq 0$ case, the calculated results reproduce the $R_{side}$ and $R_{out}$ exactly but slightly underestimate the value of $R_{long}$. Again, a reasonable parameter configuration can not be found for fitting ALICE data in the case of $\Delta\tau$ = 0. These results suggest that the system lifetime and particle emission duration should be taken into account at the same time while describing $R_{side}$, $R_{out}$ and $R_{long}$ with the same parameter configuration in the blast-wave model.

\section{Summary}

The HBT radii ($R_{side}$, $R_{out}$ and $R_{long}$) are calculated from the blast-wave model in the ``out-side-long" ($o s l$) coordinates system. In comparison with the experimental data~\cite{HBT-BES-EXP-STAR,HBT-ALICE}, we found that: in the case of $\Delta\tau\neq 0$, the parameter configuration for blast-wave model can successfully describe the experimental results of collision energy and transverse momentum dependence of $R_{side}$, $R_{out}$ and $R_{long}$. Since the collision system has different temperature at each centre-of-mass energy point, the configured parameters can be considered as the preferred values with a case of $T_{kin}$ as a function of $\sqrt{s_{NN}}$ and $\Delta\tau\neq 0$ as shown in the Figure~\ref{fig:R0} and~\ref{fig:tau0Deltatau}. However, it can not be configured for the blast-wave parameter to fit the experimental data while setting the $\Delta\tau$ to zero. This may imply that the particle emission duration plays an important role to describe the system expanding and can not be ignored while calculating the $R_{side}$, $R_{out}$ and $R_{long}$ to fit the data at the same time. And the difference of $R_{out}^2-R_{side}^2$ presents a non-monotonic increasing trend with the increasing of $\sqrt{s_{NN}}$ as seen in the experimental analysis~\cite{HBT-BES-EXP-STAR}, which is sensitive to the equation of state and might be related to the critical end point with other observables taken into account.
 
This work was supported in part by the Major State Basic Research
Development Program in China under Contract No. 2014CB845400, the
National Natural Science Foundation of China under contract Nos. 11421505,
11220101005,  11105207, 11275250, 11322547 and U1232206, and the CAS Project Grant No. QYZDJ-SSW-SLH002.

\end{document}